%% file: hollow_cathode.tex
\documentclass[%
 aip,numerical,
 amsmath,amssymb,
 reprint,%
 onecolumn
]{revtex4-1}

\input{customed_settings_pop}

\makeatletter
\def\@email#1#2{%
 \endgroup
 \patchcmd{\titleblock@produce}
  {\frontmatter@RRAPformat}
  {\frontmatter@RRAPformat{\produce@RRAP{*#1\href{mailto:#2}{#2}}}\frontmatter@RRAPformat}
  {}{}
}%
\makeatother
\begin{document}

\preprint{AIP/123-QED}

\title[Establishing Criteria for the Transition from Kinetic to Fluid Modeling in Hollow Cathode Analysis]{Establishing Criteria for the Transition from Kinetic to Fluid Modeling in Hollow Cathode Analysis}
\author{W. Villafana}
 \email{wvillafa@pppl.gov}
 \affiliation{Princeton Plasma Physics Laboratory, Princeton NJ 08543 USA}
\author{A. T. Powis}
\affiliation{Princeton Plasma Physics Laboratory, Princeton NJ 08543 USA}
\author{S. Sharma}
\affiliation{Basic Theory and Simulation Division, Institute for Plasma Research, Gandhinagar 382428, India}

\author{I.~D.~Kaganovich}%
\affiliation{Princeton Plasma Physics Laboratory, Princeton NJ 08543 USA}
\author{A.~V.~Khrabrov}
\affiliation{Princeton Plasma Physics Laboratory, Princeton NJ 08543 USA}


\date{\today}

\begin{abstract}
In this study, we conduct 2D3V Particle-In-Cell simulations of hollow cathodes, encompassing both the channel and plume region, with an emphasis on plasma switch applications. 
The plasma in the hollow cathode channel can exhibit kinetic effects depending on how fast electrons emitted from the insert are thermalized via Coulomb collisions. 
The criterion that determines whether the plasma operates in a fluid or kinetic regime is given as follows.
When Coulomb collisions occur at a much greater rate than ionization or excitation events, the Electron Energy Distribution Function relaxes to a Maxwellian distribution and the plasma within the channel can be described with a fluid model.
In contrast, if inelastic processes are much faster, then the Electron Energy Distribution Function in the channel exhibits a notable high-energy tail, and a kinetic treatment is required.
This criterion is applied to other kinds of hollow cathodes from the literature, revealing that a fluid approach is suitable for most electric propulsion applications, whereas a kinetic treatment might be necessary for plasma switches.
Additionally, a momentum balance reveals that a diffusion equation is sufficient to predict the plasma plume expansion, a crucial input in the design of hollow cathodes for plasma switch applications.
\end{abstract}

\maketitle

\section{\label{sec:level1}Introduction}

Hollow cathodes are efficient plasma sources for numerous applications, including but not limited to, spacecraft plasma propulsion \citep{goebelPlasmaHollowCathodes2021,levRecentProgressResearch2019,siegfriedModelMercuryOrificed1984,krishnanPhysicalProcessesHollow1977,taunayPhysicsThermionicOrificed2022a,taunayPhysicsThermionicOrificed2022}, beam injectors for fusion devices and instruments \citep{goebelPlasmaStudiesHollow1982,deichuliIonSourceLaB62006}, surface processing \citep{morgnerHollowCathodeHighperformance1998,kuoHotHollowCathode1986,barankovaHollowCathodeHybrid2006,childEnhancedHollowCathode2015}, plasma-material interaction studies \citep{matlockDcPlasmaSource2014}, as well as plasma switch technologies \citep{goebelLowVoltageDrop1993,goebelColdCathodePulsed1996,meshkovElectricalThermalCharacteristics2024}. 
Thermionic hollow cathodes feature an insert coated with a low-work function material such as graphite. 
Following an initial heating stage induced by an external process, such as a radio-frequency (RF) exciter or a heater, the primary emitted electrons acquire the energy required for ionization; subsequently triggering the plasma ignition.
The cathode then self-heats mainly through ion bombardment at the wall, leading to a self-sustaining discharge. 
The plasma generated thereafter reaches a high density before spreading out of the tube to produce a plume. 
Although extensively studied in the literature, hollow cathodes present a rich physics environment that is not yet fully understood.
First, the plasma density, and sometimes the neutral pressure in the case of space propulsion applications, drops by several orders of magnitude between the channel region and the plume, considerably increasing the mean free path of particles \citep{goebelPlasmaHollowCathodes2021}.
Furthermore, multiple studies have reported that the plume area could be the site of instabilities that may lead to the formation of high energy ions \citep{goebelPotentialFluctuationsEnergetic2007,dodsonIonEnergyWave2018,suzukiEnergeticIonPlasma2023}.
Finally, the design of hollow cathodes has undergone significant enhancements through a series of incremental improvements. 
These advances include the adoption of novel emitter materials \citep{levRecentProgressResearch2019,saridedeLaB6HollowCathode2023} along with the incorporation of low sputtering-yield materials and fine-tuning of the cathode geometry for optimal performance.
For instance, adjusting the size and shape of the orifice, and separating the channel chamber from the plume area can affect the emitter temperature profile and thus the discharge current \citep{guerrerovelaPlasmaSurfaceInteractions2019}.
Thanks to these innovations, both the lifespan and operational efficiency of hollow cathodes have experienced gradual and notable enhancements \citep{pedriniModelingLaB6Hollow2015,goebelPlasmaHollowCathodes2021}.
The improvement in the hollow cathode design was made possible thanks to inputs provided by numerical modeling, which became increasingly accurate over time. 
The detailed description of the plasma flow in both the channel and the plume, and the in-depth understanding of potential instabilities are crucial for future hollow cathode design.

Comprehensive modeling of hollow cathodes is challenging as the plasma may operate in different regimes.
In the channel, where the plasma density is high and assumed near equilibrium, a fluid model is often appropriate, whereas, in the plume, kinetic effects may appear. 
Historically, the channel region has attracted most of the interest, and therefore most descriptions of hollow cathodes are based on fluid models \citep{goebel2008fundamentals}. 
Zero-dimensional (0D) \citep{taunay0DModelOrificed2019,gurciulloNumericalStudyHollow2020} and 1D models \citep{katzOneDimensionalHollowCathode2003,panelliDevelopmentValidationSimplified2018} focus mainly on the inner channel and orifice area and can give quick estimates of plasma production and temperature.
A better description of plasma-wall interaction and neutral flow effects at the insert can be obtained with 2D models \citep{mikellidesHollowCathodeTheory2005,mikellidesPlasmaProcessesDispenser2006} using the Richardson-Dushman equation \citep{dushmanElectronEmissionMetals1923}.
This equation governs the electron emission and improves understanding of erosion mechanisms \citep{mikellidesWearMechanismsElectron2008}.
Capitalizing on these efforts, 2D models have been extended to the near plume region \citep{mikellidesNeutralizerHollowCathode} often including self-consistent resolution of the insert temperature \citep{saryHollowCathodeModeling2017,saryHollowCathodeModeling2017a,cong2DModelLowpressure2018}.
One motivation for including the plume area is the potential presence of anomalous resistivity leading to higher electron temperature, than what classical transport and Ohm's law would predict \citep{mikellidesEvidenceNonclassicalPlasma2007,georginExperimentalEvaluation2D2021}.
Such an effect seems to originate from ion acoustic turbulence (IAT) identified by Jorns \textit{et al.}~\citep{jornsIonAcousticTurbulence2014}, leading to the creation of energetic ions that ultimately may erode the cathode \citep{mikellidesWearMechanismsElectron2008,goebelPotentialFluctuationsEnergetic2007}.
The IAT was further investigated in \refsonlinecite{tsikataCharacterizationHollowCathode2021,haraIonKineticsNonlinear2019} in the context of hollow cathode for space propulsion. 

Fluid models account for IAT via an adjustable parameter to fit experimental measurements \citep{mikellidesNumericalSimulationsPartially2015,saryHollowCathodeModeling2017,mikellidesDischargePlasmaIon2010}.
However, since fluid models cannot self-consistent capture kinetic effects, such as the IAT, this parameter needs to be adjusted for each specific configuration using information obtained from experiments or kinetic simulations.

Efforts to address these kinetic effects can be achieved by explicitly accounting for a separate population of fast electrons in the fluid model, distinct from the near-Maxwellian low energy plasma electrons.
This approach was successfully demonstrated in \refonlinecite{bogdanovModelingShortDc2013}, which presented a model of a thermionic cathode that can serve as a current and voltage stabilizer investigated in experimental studies~\citep{mustafaevControlCurrentVoltage2012a,mustafaevSharpTransitionTwo2014}.
Additional kinetic effects may be captured by employing hybrid methods, as shown in \refsonlinecite{kubotaHybridPICSimulationPlasma2016,kubotaComparisonsHybridPICSimulation2019}.
In these studies, electrons are represented using a drift diffusion model, whereas heavy particles, such as ions and neutrals, are treated as particles tracked via a particle-in-cell (PIC) approach. 
These modeling techniques result in favorable comparisons with experimental data obtained from the LaB6 hollow cathode.
A full kinetic model of a hollow cathode for space propulsion was investigated in \refonlinecite{levkoTwodimensionalModelOrificed2013}.
In this work a 2D-2V PIC simulation modeled a miniaturized orificed hollow cathode. 
A parametric study over the cathode voltage, gas pressure and radius size was performed and subsequent effects on the steady state parameters were reported.
The simulation was sped up using a miniaturized geometry and by lowering the mass of the Xenon gas by a factor of 100.
Non-Maxwellian behavior of the Ion and Electron Velocity Distribution Functions (IVDF and EVDF) was identified in both the channel and plume area.
Two other full PIC simulations with cylindrical geometry were investigated \refsonlinecite{caoNumericalSimulationPlasma2018,caoModelingPlasmaEnergy2019}, where it was again suggested that non-Maxwellian behavior for electrons exists in the insert region.

These physics and modeling challenges are mostly described in the context of electric propulsion but also apply to hollow cathodes utilized as plasma switches, which differ significantly from those employed for space propulsion in two main aspects.
First, the neutral pressure is approximately uniform in both the channel and the plume, whereas it exhibits a strong gradient for hollow cathodes used in space propulsion. 
Second, the plume expands into closed boundaries linked to an external circuit whereas for propulsion applications the plasma expands into a vacuum.
Hollow cathodes designed for plasma switch applications represent a promising technology for the development of future power grids utilizing direct current (DC) instead of alternating current (AC) at medium or high voltage\citep{CIGRE2023HVDCBreakers,CIGRE2023MVDC,CIGRE2023HVDC}.
DC transmission is an attractive solution for meeting growing electricity demand and integrating the intermittency of renewable energies.
Indeed, it offers lower capacitive and inductive losses for a given right-of-way with respect to AC transmission lines.
In addition, it provides greater control and stability, avoiding the synchronization issues that can affect AC grids. 
However, as reported in \refonlinecite{meshkovElectricalThermalCharacteristics2024}, DC circuit breakers are inherently more costly and complex than AC breakers, as a classic mechanical switch opens at a relatively slow rate, which would result in dangerous and destructive arcing.
In this context, the authors in \refonlinecite{meshkovElectricalThermalCharacteristics2024}, experimentally explored different geometries, gases, and operating conditions of a hollow cathode that could potentially serve as a DC circuit breaker.
In particular, they noted that the discharge voltage across the cathode would drop from $\sim 30-$\SI{40}{\volt} before reaching a plateau corresponding to the ionization or excitation potential of the gas, depending on the gas pressure, which ranged from \SI{30}{\milli\torr} to \SI{1.5}{\torr}, and the discharge current.
They also reported that the ionization mechanism is likely changing with the various test conditions. 
When the voltage drop, corresponds to a lower plasma density, the ionization is mostly dominated by direct ionization from primary electrons stemming from the cathode.
In contrast, when the plasma density is high, a regime for which the voltage drop plateau has been reached, ionization is mostly due to the heating of the plasma bulk by the primary electrons.
 
We pursue two goals in the present paper. 
The first is to determine a general criterion for when either a fluid or kinetic approach is most suited to model the discharge channel. 
We then investigate how the plasma expands into the plume, which is a crucial input for plasma switch design. 
To do so, we perform kinetic modeling of a hollow cathode for plasma switch applications, which, to the best of our knowledge, has not previously been reported. 
This paper is organized as follows. 
In \Cref{sec_numerical_setup} we describe in detail the numerical setup, and discuss the main assumptions underlying our modeling. 
In \Cref{sec_transition_kinetic_fluid_regime} the plasma dynamics at steady state is elucidated, before we conduct an in-depth examination of the Electron Energy Distribution Function (EEDF) and develop our criterion for kinetic or fluid modeling.
Finally, in \Cref{sec_plume_region} the plasma plume is analyzed and estimated with a simple analytical law.

\begin{figure}[!htb] 
	\centering
	\includegraphics{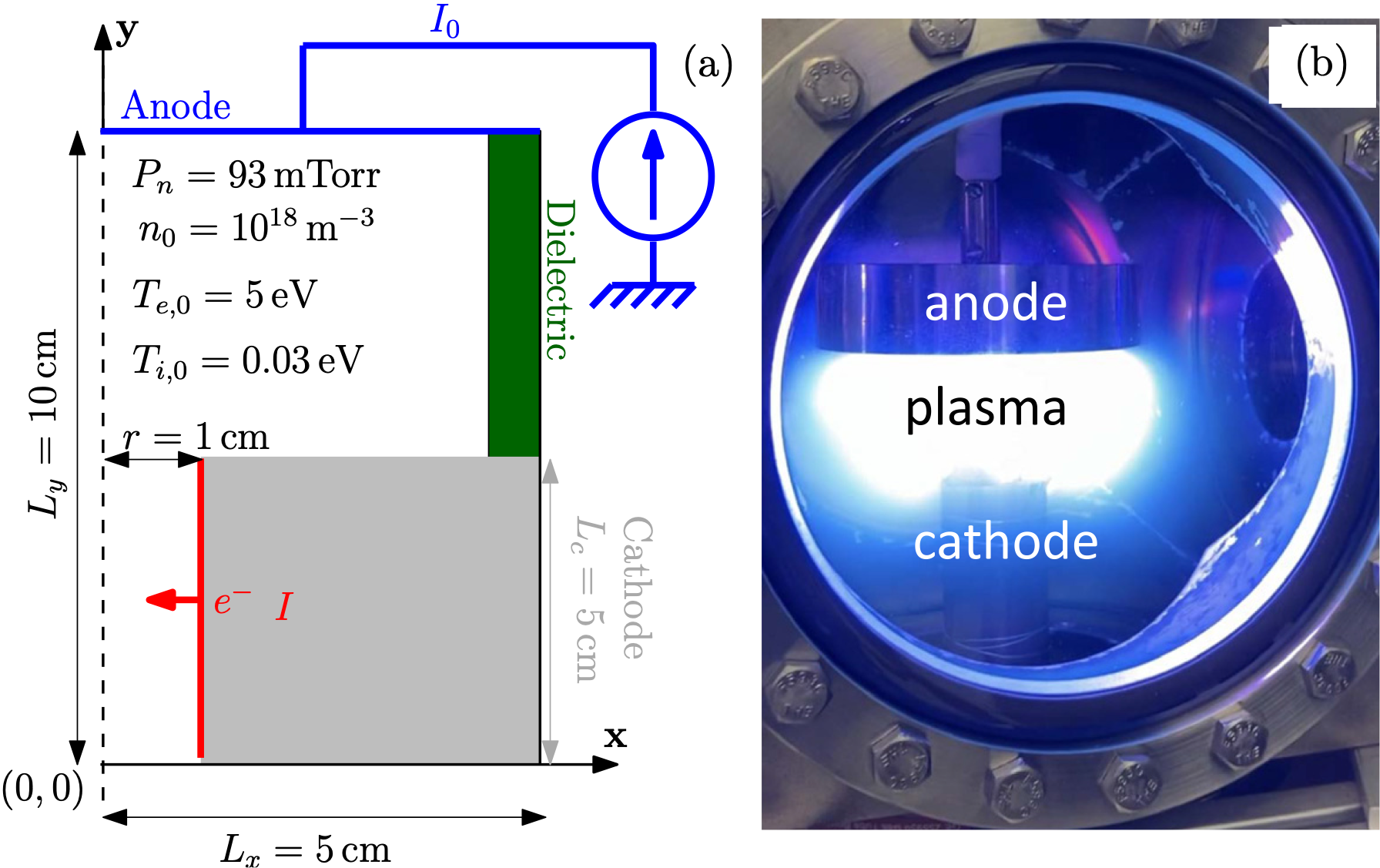}
	\caption{(a): The computational domain for the hollow cathode switch simulations. The solid dashed line on the left is the symmetry axis. The cathode, modeled as a gray block is grounded and emits electrons into the channel. The bottom of the channel is grounded and fully absorbing. The anode at the top is connected to a current source. A thin dielectric layer, purposely enlarged for clarity, separates the cathode and the anode. (b): Recent experimental setup of a hollow cathode used for plasma switch applications, reprinted from A. I. Meshkov \textit{et al.}, Physics of Plasmas \textbf{31}, 013503 (2024)~\citep{meshkovElectricalThermalCharacteristics2024}, with the permission of AIP Publishing.}
	\label{sec_2_schematic_schematic_PAPER_2023}
\end{figure}

\section{Numerical setup}
\label{sec_numerical_setup}

The configuration considered in this work is based on a hollow cathode experimental setup described in \refonlinecite{meshkovElectricalThermalCharacteristics2024}.
The computational domain is comprised of a rectangular box of size $L_x \times L_y$ with geometry as shown in \Cref{sec_2_schematic_schematic_PAPER_2023}.
The channel of the hollow cathode is located between the symmetry axis on the left of the simulation domain and the conducting cathode surface to the right, with dimensions $r \times L_c$.
A thin dielectric layer, spanning only a few cells in width, is positioned between the cathode and the anode along the right boundary (shown in green).
The anode (shown in blue) is connected to an external circuit for which a constant current, denoted as $I_0$, is applied.
The simulation is performed using the Cartesian version of the explicit momentum conserving PIC-MCC code EDIPIC-2D \citep{EDIPIC2DDocMain}, which has been verified through numerous international benchmarks \citep{charoy2DAxialazimuthalParticleincell2019,villafana2DRadialazimuthalParticleincell2021,turnerSimulationBenchmarksLowpressure2013}, and used to model a wide variety of low-temperature plasma systems \citep{xuRotatingSpokesPotential2023,sunDirectImplicitExplicit2023,sunElectronModulationalInstability2022,raufParticleincellModelingElectron2023,sonUnintendedGasBreakdowns2023}.
Although the actual hollow cathode is cylindrical, the generic criterion that determines whether the plasma operates in a fluid or kinetic regime does not depend on the geometry. 
Besides, PIC modeling in cylindrical coordinates can suffer from enhanced numerical noise near the centerline~\citep{haraEffectsMacroparticleWeighting2023} which can distort significantly the results. 
Therefore, in order to avoid unnecessary difficulties, a cylindrical modeling was not retained in this work. 

\subsection{Charged particles, species, and discretization}

The plasma consists of electrons and singly charged argon $Ar^+$ macroparticles with fixed statistical weight $p_w=\SI{2.92e7}{}$. 
A constant neutral background of $Ar$ is maintained at a pressure of $P_n=\SI{93}{\milli\torr}$, corresponding to a density of $n_n=\SI{3e21}{\per\cubic\meter}$ and temperature of \SI{300}{\kelvin}. 
We reiterate that this represents a major difference from hollow cathodes used in electric propulsion for which the neutral pressure exponentially decreases from the channel exit into the vacuum.
Ions can only interact with neutrals via charge exchange collisions \cite{maiorovIonDriftGas2009}, whereas electrons collide with the neutral background via elastic, excitation, or ionization collisions using state-of-the-art cross-section data \citep{pancheshnyiLXCatProjectElectron2012}.
Due to high electron densities, Coulomb collisions between electrons are also implemented via Nanbu's approach \citep{nanbuTheoryCumulativeSmallangle1997}. 
This was not done in previous fully PIC modeling of hollow cathode to our knowledge. 
No recombination reactions are considered in the present setup.
Neutral metastables are not modeled in this work and so step-wise ionization is not considered.
For space propulsion applications, step-wise ionization is usually omitted as it does not seem to affect plasma production significantly~\citep{saryHollowCathodeModeling2017}. 
For plasma switch applications, in large cathodes~\citep{meshkovElectricalThermalCharacteristics2024}, the presence of metastables might lower the plasma potential in the bulk as step-wise ionization requires a lower acceleration of electrons emitted from the cathode to effectively produce plasma. 
Although interesting, we reserve this work for a companion paper as it is out of the scope of the present study. 

The simulation is initialized with a uniform plasma density of $n_0=\SI{e18}{\per\cubic\meter}$ and a Maxwellian distribution with a temperature of $T_{e,0}=\SI{5}{\electronvolt}$ and $T_{i,0}=\SI{0.03}{\electronvolt}$ for electrons and ions, respectively. 
The cell size $\Delta x$ and time step $\Delta t$ are chosen such that they satisfy the accuracy and stability conditions for an explicit momentum conserving PIC scheme \citep{birdsallPlasmaPhysicsComputer1985}, i.e., $\Delta x<\lambda_{De}$ and $\Delta t<0.2\omega_{pe}^{-1}$, where $\lambda_{De}$ and the $\omega_{pe}$ are the Debye length and the electron plasma frequency respectively. 
In the present simulations, the maximum density is expected to reach approximately $\sim$\SI{5e19}{\per\cubic\meter}, which with an electron temperature of a few electronvolts, would require a cell size of a few microns and a time step less than a picosecond.
The subsequent computational cost would be substantial for the considered computational domain and a simulation time of over a few hundred of microseconds. 
To mitigate this cost, the vacuum permittivity is scaled by a factor $\epsilon_r \approx 1057$, which increases the allowed time step and cell size by a factor $\sim$ 33.
Such a value was chosen to enlarge the cells enough to get dimensions similar to what is used in \refonlinecite{meshkovElectricalThermalCharacteristics2024}.
Such techniques have been widely used in low-temperature plasma simulations in particular for space propulsion \citep{taccognaPlasmaPropulsionModeling2023,taccognaThreedimensionalParticleincellModel2018} and also for hollow cathodes \citep{caoNumericalSimulationPlasma2018,caoModelingPlasmaEnergy2019}.
In the latter two PIC studies, it was shown that scaling $\epsilon_r$ by a factor of $900$, did not significantly affect plasma bulk properties, which guided our choice of $\epsilon_r$.
The chosen time step and cell size are $\Delta t=\SI{12.1}{\pico\second}$ and $\Delta x=\SI{44}{\micro\meter}$, respectively.
In the most constrained scenario, utilizing the electron and temperature density from the channel, we obtain $0.2/\omega_{pe}\approx\SI{37}{\pico\second}$ and $\lambda_{De}\approx\SI{100}{\micro\meter}$ guaranteeing a high spatio-temporal resolution for the simulation. 

\begin{table}
	\begin{center}
    	\ra{1.3} 
    	\begin{tabular}{@{}  l l l l @{}}
    	
        	\toprule
        	Parameters & \text{Symbol} & Value & Unit  \\
        	\midrule
        	\multicolumn{4}{c}{\textbf{Simulation domain}}  \\
        	\cmidrule{1-4}
        	Cell size & $\Delta x$ & \SI{44}{} &\SI{}{\micro\meter}   \\

			Number of cells  & $N_{cell}$ & 2273x1153 \\
        	
        	Radial length  & $L_x$ & 5 &\SI{}{\centi\meter}  \\
        	
        	Axial length  & $L_y$ & 10 &\SI{}{\centi\meter}  \\ 	

        	Channel length  & $L_c$ & 5 &\SI{}{\centi\meter}  \\ 	
        	
        	Channel radius  & $r$ & 1 &\SI{}{\centi\meter}  \\ 	        	
			
        	\midrule
        	\multicolumn{4}{c}{\textbf{Operating conditions}}  \\
        	\cmidrule{1-4}
        	
        	External circuit current density & $J_0$ & \SI{2.06}{} &\SI{}{\kilo\ampere\per\meter\squared} \\

        	Thermionic current & $J$ & \SI{2.06}{} &\SI{}{\kilo\ampere\per\meter\squared} \\

        	Gas pressure & $P_n$ & \SI{93}{} &\SI{}{\milli\torr} \\
        	
        	Injection temperature & $T_{inj,e}$ & \SI{0.15}{} &\SI{}{\electronvolt} \\        	        
        	
        	\midrule
        	\multicolumn{4}{c}{\textbf{Computational parameters}}  \\
        	\cmidrule{1-4}

        	Time step & $\Delta t$ & \SI{12.1}{} &\SI{}{\pico\second} \\     
        	
        	Permittivity scaling factor & $\epsilon_r$ & \SI{1057}{} &\SI{}{} \\             	   
      	        	        	
        	\bottomrule
	    \end{tabular}
	\end{center}
    \caption{PIC simulation parameters for hollow cathode switch simulation reference case C0.}
    \label{Table_numerical_setup}	
\end{table}

\subsection{Field solver and boundary conditions}

The system is symmetric with respect to a centerline at $x=\SI{0}{\centi\meter}$ with Figure \Cref{sec_2_schematic_schematic_PAPER_2023} showing the right half of the cathode and plume regions modeled in this paper. 
A Neumann boundary condition for the field solver is imposed at the symmetry axis while particles crossing $x=\SI{0}{\centi\meter}$ are reflected back into the domain. 
The cathode, whose inner radius is \SI{1}{\centi\meter} is modeled as a conducting and grounded material that fully absorbs particles hitting its surface.
The cathode also emits thermionic \SI{0.15}{\electronvolt} electrons from the surface into the channel region. 
In contrast to other studies that rely on the Richardson-Dushman law \citep{dushmanElectronEmissionMetals1923} to determine the electron flux emitted by the cathode wall, the flux is determined by a preset fixed current $I$.
The anode, located at $y=L_y$, is a fully conducting surface connected to an external circuit comprised of a current source $I_0$ and the electric potential at the anode automatically adjusts to ensure the current received from the plasma is equal to $I_0$.
The implemented external circuit model is described in the work by Vahedi and DiPeso \citep{vahediSimultaneousPotentialCircuit1997}.
A thin dielectric layer, enlarged for clarity in \Cref{sec_2_schematic_schematic_PAPER_2023}, connects the anode and the cathode, similarl to the configuration of experimental plasma switches \citep{goebelLowVoltageDrop1993}. 
Surface charge accumulating on the dielectric prevents a net current flow from leaking to the side at $x=L_x$.
The plasma potential is obtained self consistently by solving the 5-point discrete Poisson's equation using the PETSc linear algebra package \citep{balayPETScWebPage2023} with the generalized minimal residual (GMRES) solver.
The electric field is then calculated from the potential using a second-order finite volume method.

In \Cref{Table_numerical_setup}, the main numerical parameters are reported for our reference case C0.
Values of currents $I$ and $I_0$ are chosen such that the corresponding current densities, $J$ and $J_0$, computed in the present Cartesian geometry, are similar to those expected in a cylindrical hollow cathode for plasma switch application. 
$J$ and $J_0$ are computed by dividing the currents by the area $A=L_c\times\SI{1}{\meter}$ as the off-plane direction defaults to one meter length.
Other cases, labeled from C1 to C5, with different values for the external circuit and emitted current densities, $J_0$ and $J$, will be considered in \Cref{Subsec_parametric_study}.

\subsection{Assessment of numerical thermalization}

When studying energy distribution functions within (PIC) simulations, it is crucial to discern real physical phenomena from artifacts such as numerical thermalization.
Recent findings, notably in \refonlinecite{jubinNumericalThermalization2D2024}, highlight that numerical thermalization can be a prevalent issue in PIC simulations.
This phenomenon may inadvertently skew velocity distributions towards a Maxwellian profile at an artificially high rate, depending on the chosen numerical parameters. 
Since the main findings of our work rely heavily on measures of thermalization, it is critical to assess the impact of this numerical effects for our numerical parameters.
From the foundational work of Hockney~\citep{hockneyMeasurementsCollisionHeating1971}, the thermalization time $\tau_{therm}$ can be estimated using,

\begin{equation}
	\tau_{therm} = \frac{2\pi}{\omega_{pe}}\frac{N_{ppc}}{0.98}\left(1+\left(\frac{\lambda_D}{\Delta x}\right)^2\right),
	\label{Eq_hockney}
\end{equation}

\noindent where in our simulations $N_{ppc} \approx 700$ represents the number of particles per cell in the channel, $\omega_{pe}=\SI{5.20e9}{\radian\per\second}$ is the plasma frequency, $\Delta x = \SI{44.2}{\micro\meter}$ is the grid spacing and $\lambda_D = \SI{128}{\micro\meter}$ the Debye length in the channel.
This yields $\tau_{therm}\approx\SI{6.90}{\micro\second}$, which, when compared with the thermalization time due to Coulomb collisions, $\tau_{coulomb}=\nu_{ee}^{-1}\approx\SI{0.10}{\micro\second}$, is 20 times higher.
Consequently, in our simulation setup, numerical thermalization has no significant influence on the results.
This outcome is primarily attributed to the implementation of a scaled relative permittivity, which \refonlinecite{jubinNumericalThermalization2D2024} identified as a mechanism to reduce the effect of numerical thermalization.

\section{Transition from the kinetic to fluid regime in the channel} 
\label{sec_transition_kinetic_fluid_regime}
\subsection{Steady state operation of the hollow cathode plasma switch}

\begin{figure}[!htb] 
	\centering
	\includegraphics{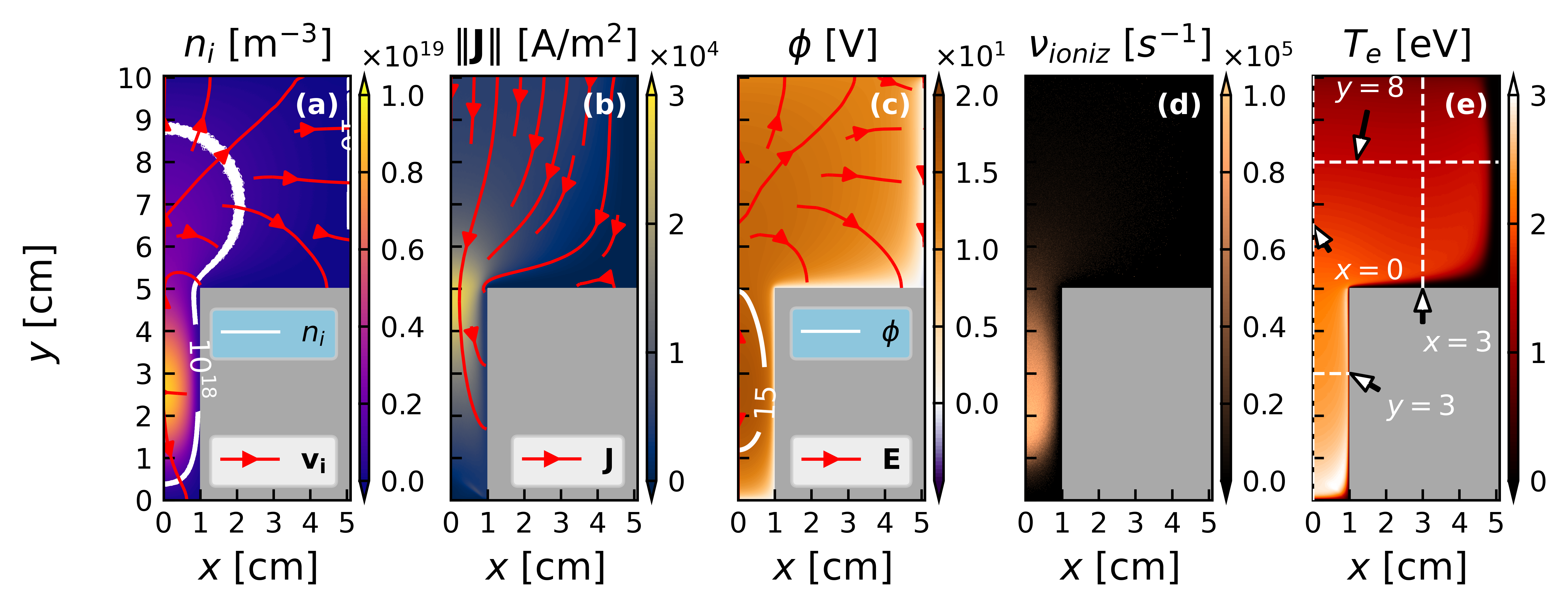}
	\caption{2D maps of physical quantities at steady state for the C0 case: (a) ion density $n_i$, (b), current density magnitude $\Vert \mathbf{J} \Vert$, (c) plasma potential $\phi$, (d) ionization collision frequency $\nu_{ioniz}$, (e) total electron temperature $T_e$. The total electron temperature is defined as the average of the temperatures in the $x$, $y$ and $z$ directions, respectively.  Streamlines in (a-b-c) are ion velocity $\mathbf{v_i}$, total current density $\mathbf{J}$ and electric field $\mathbf{E}$. In (a), iso-contours of the ion density $n_i$ are also shown. In (e), the white dashed lines indicate where 1D cuts are taken in \Cref{Fig_1D_cuts_general}.} 
	\label{Fig_results_grid_plot}
\end{figure}

\begin{figure}[!htb] 
	\centering
	\includegraphics{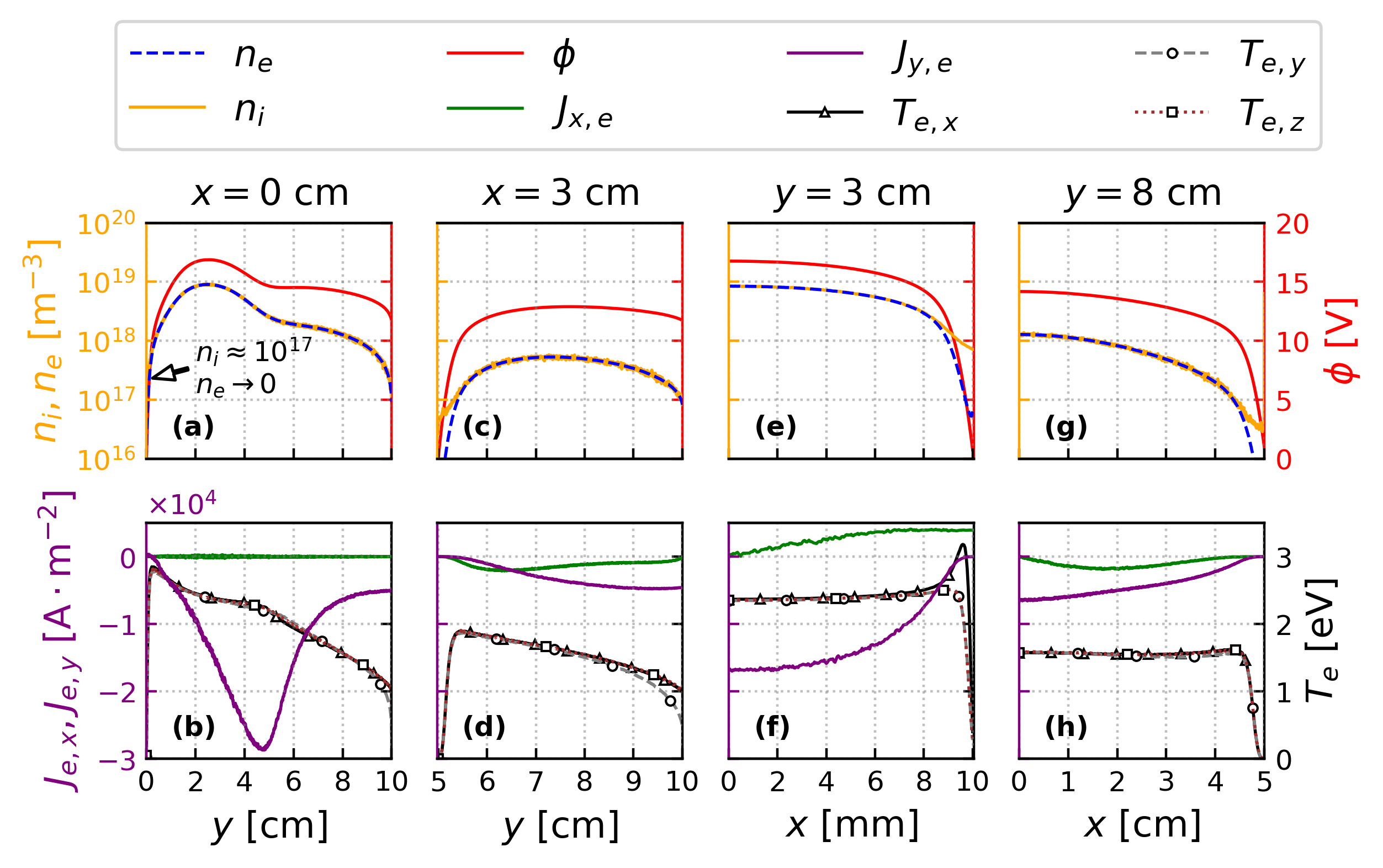}
	\caption{1D cuts of the main physical quantities at steady state in the channel and plume. Cuts are represented as white dashed lines in \Cref{Fig_results_grid_plot} (e). Cuts at $x=0$, $x=3$, $y=3$ and $y=8$ are sorted by column. At the top row the ion density $n_i$, the electron density $n_e$, and the plasma potential $\phi$ are illustrated for the four cuts. At the bottom row, the electron current density $J_{e,x}$ and $J_{e,y}$ in the $x$ and $y$ directions, as well as the three electron temperature components $T_{e,x}$, $T_{e,y}$ and $T_{e,z}$ are depicted for the four cuts. Part (d) and (f) show a temperature anisotropy near the the anode and cathode wall, respectively. Parts (a) and (b) show the absence of a sheath neat the anode wall.}
	\label{Fig_1D_cuts_general}
\end{figure}

\Cref{Fig_results_grid_plot} shows time averaged 2D contour plots of the hollow cathode switch operating at steady state. 
Electrons are created at the thermionic surface of the cathode and then accelerate through the $\sim16-\SI{17}{\volt}$ potential increase from the grounded wall to the centerline, as shown in \Cref{Fig_results_grid_plot} (c). 
The energy gain is slightly above the ionization potential for Argon, \SI{15.76}{\electronvolt}, and so each injected electron may have the opportunity to ionize a single neutral, leading to frequent ionization events in the channel area as illustrated by \Cref{Fig_results_grid_plot} (d).
Similarly to what was observed in plasma switch experiments detailed in \refonlinecite{meshkovElectricalThermalCharacteristics2024}, this results in a dense diffuse glow discharge plasma in the channel, which, in the present work, leads to a maximum of density $\sim\SI{9e18}{\per\cubic\meter}$ in the plasma bulk at $x=0$, $y=2.5$ (\Cref{Fig_results_grid_plot} (a)).
The plasma density drops near the non-emissive lower boundary as the potential monotonically goes to zero, which is the behavior typically observed in hollow cathode experiments and simulations \citep{mikellidesNumericalSimulationsPartially2015,goebelPlasmaHollowCathodes2021}.

Electrons from the dense plasma in the channel expand into the plume under the effects of the pressure gradient (see \Cref{sec_plume_region} for further details). 
As they accelerate through the narrow channel exit, the total current density $J$, driven primarily by electrons, dramatically increases, reaching a maximum of $\sim$25-\SI{30}{\kilo\ampere\per\meter\squared} (\Cref{Fig_results_grid_plot} (b)). 
As expected for hollow cathode behavior, electrons are effectively extracted from the channel and eventually travel to the anode where they are absorbed.
Overall the plasma density within the plume reduces to one-tenth of the levels found in the channel area.
This density drop, from $\sim$\SI{e19}{\per\cubic\meter} to $\sim$\SI{e18}{\per\cubic\meter} is significantly lower than that observed in hollow cathode used for electric propulsion.
Indeed for space propulsion, the plasma expansion into the vacuum leads to density drops from $\sim$\SI{e21}{\per\cubic\meter} to $\sim$\SI{e17}{\per\cubic\meter}~\citep{goebelPlasmaHollowCathodes2021,saryHollowCathodeModeling2017}. 
The plasma expansion is correlated to a slight drop in the total electron temperature as shown in \Cref{Fig_results_grid_plot} (e). Here, total electron temperature is defined with respect to the pressure tensor and as the sum of the three directional components.  
Cooling of the plasma in this region is consistent with hollow cathodes used for electric propulsion\citep{goebelPlasmaHollowCathodes2021,saryHollowCathodeModeling2017}.

\Cref{Fig_1D_cuts_general}, which shows cross-sections of the data from \Cref{Fig_results_grid_plot}, allows us to scrutinize the discharge behavior more closely.
The cathode sheath, which is clearly visible in \Cref{Fig_1D_cuts_general} (e) leads to acceleration of thermionic electrons and a corresponding increase in $T_{e,x}$ as shown in \Cref{Fig_1D_cuts_general} (f).
This temperature anisotropy is then relaxed by coulomb collisions as the electrons propagate into the plasma bulk.
These effects cannot be captured by fluid codes which assume quasi-neutrality all the way up to the cathode wall~\citep{goebelPlasmaHollowCathodes2021,mikellidesNumericalSimulationsPartially2015,saryHollowCathodeModeling2017}.
Although self-consistent modeling of the sheath can be achieved with multi-fluid moment models~\citep{joncquieresFluidFormalismLowtemperature2020,sahuFullFluidMoment2020}, such models should also account for an anisotropic pressure tensor~\citep{shumlakAdvancedPhysicsCalculations2011,hakimExtendedMHDModelling2008}, in order to accurately reproduce this behavior.
Nonetheless, no fluid models are able to fully capture the non-Maxwellian nature of the EEDF, which as will be shown in \Cref{sub_section_eedf_one_case}, is responsible for the anisotropy.


Previous PIC modeling of hollow cathodes have not specifically addressed this temperature anisotropy, since although non-Maxwellian behavior is mentioned, no temperature plots are shown.
In the channel, authors in \refonlinecite{levkoTwodimensionalModelOrificed2013} showed energy distribution functions, although these were limited to less than \SI{20}{\electronvolt}. 
Since they observed a potential drop between the cathode and plasma bulk greater than \SI{20}{\electronvolt}, the beam electrons, and associated anisotropy, are not represented.
In \refsonlinecite{caoNumericalSimulationPlasma2018,caoModelingPlasmaEnergy2019}, it was shown an EEDF in the plasma bulk, however this is limited to below \SI{7}{\electronvolt} whereas the potential drop is reported as $\sim\SI{25}{\electronvolt}$.
Again these limitations prevent representing potential beam electrons present in the system.
For this reason, we suspect that anisotropy in the EEDF and electron temperature are present but not depicted in these simulations.
In these PIC studies, it is important to note that electron-electron Coulomb collisions were also not accounted for, which is the only physical mechanism able to thermalize the beam electrons. 
In our own simulations, failure to include Coulomb collisions significantly affected the plasma topology in both the plume and channel, resulting in unphysical highly non-Maxwellian EEDFs.

In the anode region, the external circuit effectively adjusts the potential drop between the plasma and the surface.
By ensuring that the electron current arriving at the anode matches the source current strength $I_0$, the potential drop is kept moderate to prevent excessive repulsion of electrons.
Consequently, a maximum potential drop of \SI{3}{\volt} is observed in parts (a) and (c) of \Cref{Fig_1D_cuts_general}, which is insufficient to break local quasi-neutrality.
Since a significant portion of electrons with moderate or high energy are allowed to escape at the anode, the Electron Velocity Distribution Function is preferentially depleted in the orthogonal direction to the anode wall ($y$-direction).
Thus, $T_{e,y}$ diverges from the two other temperature components as the plasma flow approaches the anode as shown in parts (b) and (d) of \Cref{Fig_1D_cuts_general}. 
This represents a second instance of temperature anisotropy observed in the domain.

As the plasma expands into the plume, an ambipolar electric field is created to ensure quasi-neutrality while the electrons accelerate under pressure gradient effects.
Thus, a slight decrease in plasma potential less than \SI{2}{\volt} is visible in \Cref{Fig_1D_cuts_general} (a). 
This modest voltage drop across between the channel and the anode was also reported in plasma switch experiments in \refonlinecite{meshkovElectricalThermalCharacteristics2024}, but differs from what is observed in hollow cathodes used in electric propulsion for which experiments and simulations show an increase in plasma potential between the channel and the plume~\citep{saryHollowCathodeModeling2017,saryHollowCathodeModeling2017a,goebelPlasmaHollowCathodes2021}.
Indeed, in these cases, the electrons can reach a sufficient velocity~\citep{jornsIonAcousticTurbulence2014,haraIonKineticsNonlinear2019} to trigger ion acoustic instabilities, elevating the now oscillating plasma potential in the plume.
These effects are not observed in the present work, although further discussion will be provided in \Cref{sec_plume_region}.

For completeness, the ion and electron densities, plasma potential, total current density, and the components of electron temperatures are shown at $y=\SI{8}{\centi\meter}$ in \Cref{Fig_1D_cuts_general} (g) and (h).

\subsection{Examination of the non-Maxwellian Energy Distribution Functions in the channel}
\label{sub_section_eedf_one_case}



\begin{figure}[!htb] 
	\centering
	\includegraphics{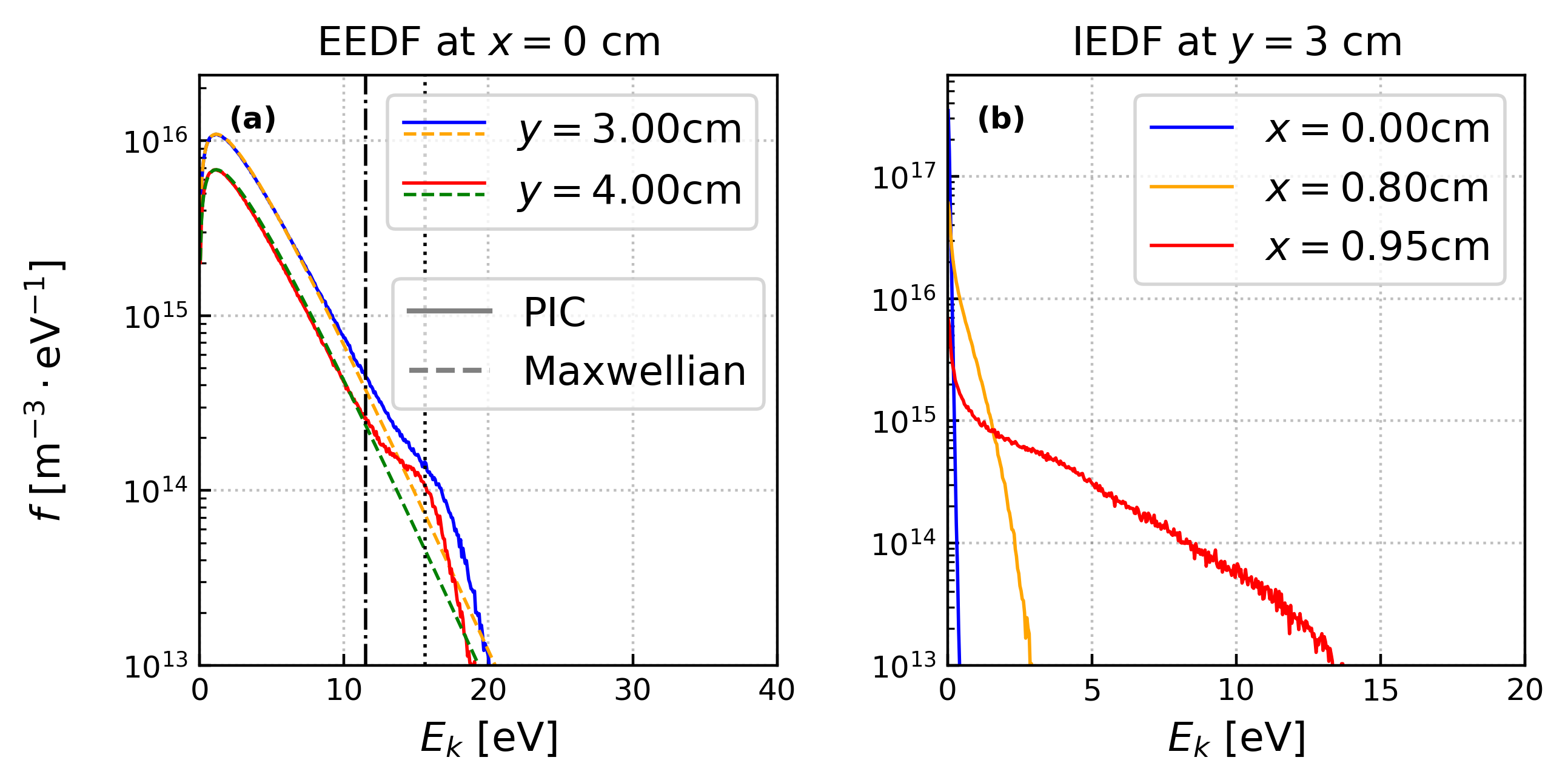}
	\caption{Energy distribution function of (a) electrons at $y=3$ cm and \SI{4}{\centi\meter} at the centerline, and of (b) ions at $x=0$ cm, 0.8 cm and \SI{0.95}{\centi\meter} at $y=\SI{3.5}{\centi\meter}$ for reference case C0. In (a), the dashed lines represent the Maxwellian EEDFs that would best fit the low energy part of the PIC measurements. The dotted and dash-dotted lines are the ionization and excitation energy potential for Argon, respectively. The EEDF exhibits a high energy tail corresponding to beam electrons and the IEDF shows the acceleration of ions as they get closer to the sheath.}
	\label{Fig_sec3_eedf_one_case}
\end{figure}

In \Cref{Fig_sec3_eedf_one_case} (a), the Electron Energy Distribution Function (EEDF) at the channel centerline  $x=0$ and axial positions $y=3$ cm and $y=4$ cm, shows two populations of electrons.
In both cases, the first population comprises of cold electrons which follow a Maxwellian distribution with a temperature of $T_e=\SI{2.4}{\electronvolt}$ given by \Cref{Eq_eedf},

\begin{equation}
	f_{Maxwellian}(E_k)  = n_{e,c}\frac{2}{\sqrt{\pi}} \sqrt{E_k} \left( \frac{1}{e T_e} \right)^{3/2} \exp\left( -\frac{E_k}{e T_e} \right),
	\label{Eq_eedf}
\end{equation}

\noindent where $n_{e,c}$ represents the density of cold electrons. 
The second population comprises of fast electrons with an energy around $16-\SI{17}{\electronvolt}$, which are clearly those emitted at the cathode and accelerated through the potential drop in the sheath. 
Estimating the Coulomb collision frequency of this population is performed via the following equation \citep{hubaNRLPlasmaFormulary1998},
\begin{equation}
	\nu_{ee} = \SI{7.7e-6}{}\frac{n_{e,c}\Lambda}{E_f^{3/2}},
	\label{eq_coulomb_nu}
\end{equation}
where $\Lambda$ is the Coulomb logarithm, $n_{e,c}$ is the density of the cold electrons background, and $E_f$ the energy of the the fast electrons. 
Electrons with energy $\SI{17}{\electronvolt}$ produce a Coulomb collision frequency $\nu_{ee}\sim\SI{9.9e6}{\per\second}$ whereas ionization and excitation collisions with the background gas have frequencies of $\nu_{ioniz}\sim\SI{1.1e7}{\per\second}$ and $\nu_{excit}\sim\SI{3.9e7}{\per\second}$ respectively. 
Therefore, fast electrons are unable to sufficiently thermalize prior to undergoing excitation or ionization collision events, meaning that the effects of this high-energy tail persist into the plasma bulk. 
As a consequence, ionization processes are mostly driven by fast electrons and an accurate assessment of ionization therefore requires a kinetic treatment.
Electrons stemming from a first excitation event, are expected to have an energy of $\sim\SI{5.5}{\electronvolt}$, which is too low for ionization.
However, Coulomb collisions of such particles with the background of cold electrons occur with a frequency of $\SI{5.37e7}{\electronvolt}>\nu_{excit}$, meaning that such electrons are likely to thermalize shortly after excitation.

\Cref{Fig_sec3_eedf_one_case} (b) shows the Ion Energy Distribution Function (IEDF) in the channel at different $x$ locations, from the centerline up to near the cathode wall. 
In the plasma bulk, at $x=0$ cm, ions follow a Maxwellian distribution.
As ions approach the sheath, starting at $x\sim 0.85$ cm the IEDF becomes skewed due to ion acceleration (\Cref{Fig_1D_cuts_general} (e) at $x=0.80$ cm and $x=0.95$ cm).
Ions accelerate under the potential drop between the plasma bulk and the sheath with a theoretical terminal energy of $\sim 16-\SI{17}{\electronvolt}$ (not depicted).



\subsection{Discussion of kinetic effects under different hollow cathode operating conditions}
\label{Subsec_parametric_study}

Coulomb collisions, responsible for plasma Maxwellization, compete with inelastic processes, ionization, and excitation.
Understanding when one is predominant over another is crucial to properly assess energy-related processes such as plasma production.
Thus, our study was extended to various hollow cathode operating conditions to identify which regimes require a kinetic treatment or when a fluid approximation may suffice, specifically we investigate how the change in potential drop from the channel to the plume influences behavior.
According to \Cref{eq_coulomb_nu}, the cold electron plasma density, determined by the injection current $I$, and the energy of fast particles, governed by the potential drop at the cathode wall are key parameters in this study.
The potential drop can be tuned by the external current $I_0$ set in the external circuit.
Indeed, whenever $I_0>I$, the hollow cathode must produce sufficient plasma to satisfy the global balance $I_0=I+I_{ioniz}-I_{loss}$, where $I_{ioniz}$ and $I_{loss}$ are the current due to ionization and wall losses respectively. 
An increase in plasma potential elevates the energy levels of emitted electrons and simultaneously enhances the confinement of cold electrons within the channel. 
This enhancement effectively raises $I_{ioniz}-I_{loss}$, thereby supplying adequate current to the external circuit. The span of operating currents and associated potential drops for our different simulation cases are provided in \Cref{Table_parametric_study}.

\begin{table}[!htb]
	\begin{center}
    	\ra{1.3} 
    	\begin{tabular}{@{}  l l l l @{}}
    	
        	\toprule
			\specialcellLeft{\textbf{Cases}\\ \ } & \specialcellLeft{Emitted current density\\ $J$ [\SI{}{\kilo\ampere\per\meter\squared}]}  & \specialcellLeft{External circuit current density\\ $J_0$ [\SI{}{\kilo\ampere\per\meter\squared}]} & \specialcellLeft{Potential difference between plasma bulk and cathode wall\\ $\Delta \phi_s$ [\SI{}{\volt}] } \\
        	\midrule

				C0 & 4.12  & 4.12  & 16.6 \\
				C1 & 2.06  & 2.06  & 16.7 \\
				C2 & 16.48 & 16.48 & 16.0 \\
				C3 & 2.06  & 2.60  & 20.0 \\
				C4 & 2.06  & 3.20  & 36.0 \\
				C5 & 80    & 80    & 15.8 \\ 	   
      	        	        	
        	\bottomrule
	    \end{tabular}
	\end{center}
    \caption{Parametric analysis of varying current density values injected at the cathode and external circuit. The reference case is labeled C0.}
    \label{Table_parametric_study}	
\end{table}

\begin{figure}[!htb] 
	\centering
	\includegraphics{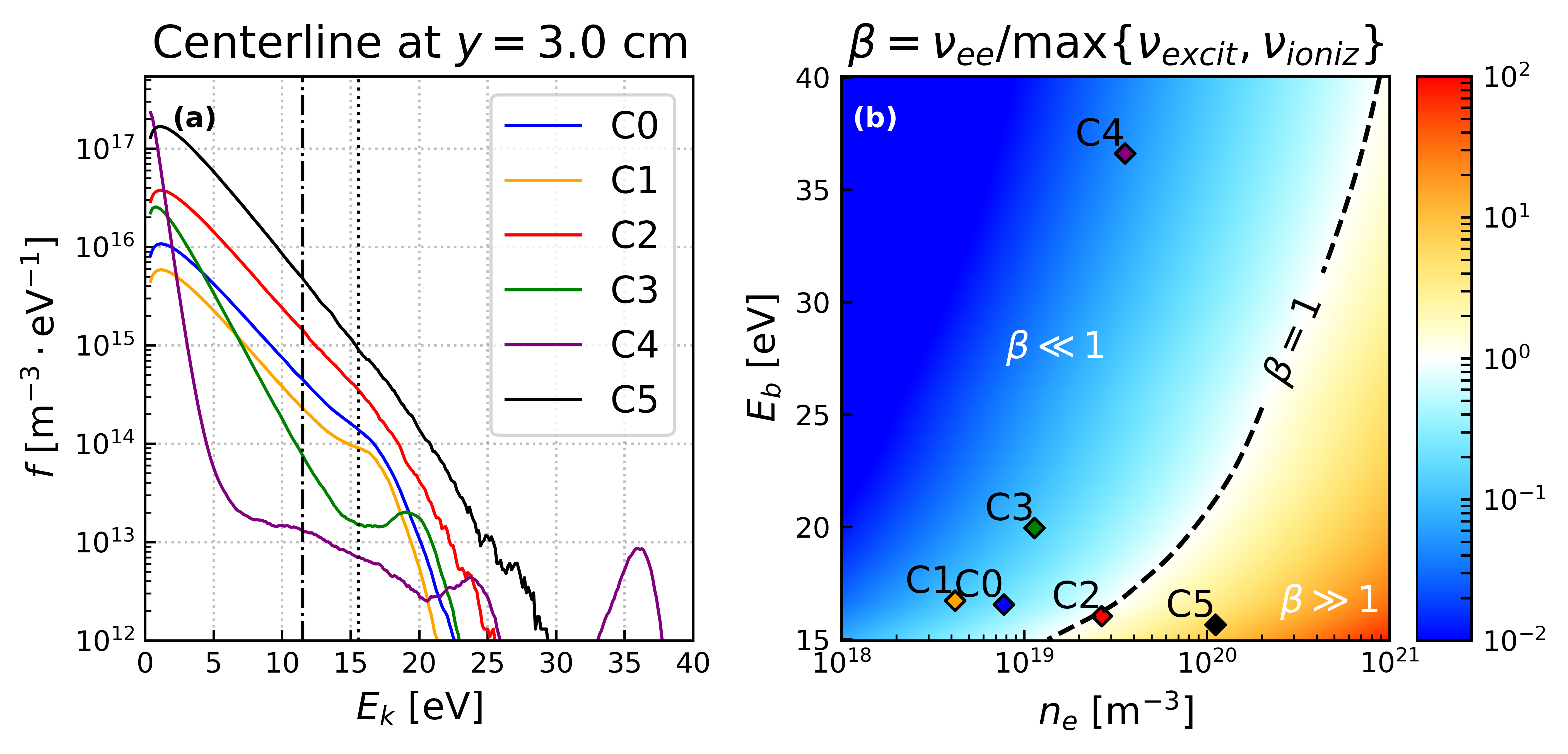}
	\caption{Assessment of kinetic effects in the hollow cathode under different operating conditions. (a): Electron Energy Distribution Function at the centerline at $y=\SI{3}{\centi\meter}$ in the channel for cases from C0 to C5 as described in \Cref{Table_parametric_study}. The dotted and dash-dotted lines are the ionization and excitation energy potential for Argon, respectively. (b): estimation of the ratio $\beta=\nu_{ee}/\max{\{\nu_{ioniz},\nu_{excit}\}}$, for various electron beam energy levels and different plasma bulk densities for an Argon gas at $P_0=\SI{93}{\milli\torr}$. Diamond symbols denote operating conditions for cases from C0 to C5.}
	\label{Fig_sec4_kinetic_effects}
\end{figure}

\noindent The EEDFs for all cases within the bulk of the channel plasma, measured at the centerline $x=\SI{0}{\centi\meter}$ and $y=\SI{3}{\centi\meter}$, are depicted in \Cref{Fig_sec4_kinetic_effects} (a).
Cases C0, C1, C2, and C5 maintain the equality $J=J_0$, while increasing the discharge current that effectively controls the plasma density level in the channel.
The presence of fast electrons is more pronounced in case C1, for which the emitted current, and the plasma bulk density, have been halved ($J_0=\SI{2.06}{\kilo\ampere\per\meter\squared}$) with respect to case C0 ($J_0=\SI{4.12}{\kilo\ampere\per\meter\squared}$).
This trend correlates with a decrease in the frequency of Coulomb electron-electron collisions, attributable to the diminished electron density.
In contrast, cases C2 and C5 present an electron density approximately three ($n_e\approx \SI{2.65e19}{\per\cubic\meter}$) and twenty times higher ($n_e\approx \SI{1.1e20}{\per\cubic\meter}$) than case C0 respectively. 
The resulting EEDF is closer to a Maxwellian as beam electrons thermalize more rapidly.
In cases C3 and C4, the current density $J$ is unchanged, however the external circuit current density $J_0$ is increased by 30\% and 60\%, respectively. 
In such cases, the beam electrons, accelerated through the cathode sheath, are fast enough to reach the center of the channel before thermalizing via Coulomb collisions. 
Thus a clear local maximum in the EEDF is visible for case C3 at an energy close to the cathode voltage drop, i.e. $\sim\SI{20}{\volt}$. 
In case C4, the scenario is more extreme with the measured cathode voltage drop approaching approximately \SI{36}{\volt}, and a corresponding EEDF maximum is present at $\sim$\SI{36}{\electronvolt}.
At this energy, electrons can ionize and excite neutral particles several times before joining the cold population, with various collision pathways shown in \Cref{Table_reactions_C4}.
A secondary local peak, observed at approximately \SI{24.5}{\electronvolt}, can be attributed to \SI{36}{\electronvolt} beam electrons that have undergone an excitation collision, considering that the excitation energy for Argon is \SI{11.5}{\electronvolt}.
This is referred as process P1 in  \Cref{Table_reactions_C4}.
Between 5 and \SI{20.3}{\electronvolt} the electron population decreases exponentially with higher energies. 
For these electrons, additional excitation of ionization processes are possible depending on their energy.
For instance, \SI{20.3}{\electronvolt} electrons are produced via process P2 and stem from \SI{36}{\electronvolt} beam electrons that have ionized once, suffering a loss of \SI{15.7}{\electronvolt}.
Similarly, \SI{24.5}{\electronvolt} electrons lead to the production of \SI{13}{\electronvolt} and \SI{8.8}{\electronvolt} electrons via excitation (process P3) and ionization (process P4), respectively. Electrons with energy
\SI{20}{\electronvolt} can ionize or excite (processes P5 and P6, respectively) one last time while \SI{13}{\electronvolt} electrons can participate in one final excitation event, denoted as process P7 in \Cref{Table_reactions_C4}.

\begin{table}[!htb]
	\begin{center}
    	\ra{1.3} 
    	\begin{tabular}{@{} l l @{}}
    	
        	\toprule
        	\textbf{Process} &  \\
        	\midrule

				P1 & $\mathrm{e}^{-1}$ (\SI{36}{\electronvolt})  + Ar $\xrightarrow{\Delta = -\SI{11.5}{\electronvolt}}$ $\mathrm{e}^{-1}$ (\SI{24.5}{\electronvolt}) + Ar  \\
				P2 & $\mathrm{e}^{-1}$ (\SI{36}{\electronvolt})  + Ar $\xrightarrow{\Delta = -\SI{15.7}{\electronvolt}}$ $\mathrm{e}^{-1}$ (\SI{20.3}{\electronvolt}) + Ar  \\
				P3 & $\mathrm{e}^{-1}$ (\SI{24.5}{\electronvolt})  + Ar $\xrightarrow{\Delta = -\SI{11.5}{\electronvolt}}$ $\mathrm{e}^{-1}$ (\SI{13}{\electronvolt}) + Ar  \\
				P4 & $\mathrm{e}^{-1}$ (\SI{24.5}{\electronvolt})  + Ar $\xrightarrow{\Delta = -\SI{15.7}{\electronvolt}}$ $\mathrm{e}^{-1}$ (\SI{8.8}{\electronvolt}) + Ar  \\    
				P5 & $\mathrm{e}^{-1}$ (\SI{20.3}{\electronvolt})  + Ar $\xrightarrow{\Delta = -\SI{15.7}{\electronvolt}}$ $\mathrm{e}^{-1}$ (\SI{4.6}{\electronvolt}) + Ar  \\  
				P6 & $\mathrm{e}^{-1}$ (\SI{20.3}{\electronvolt})  + Ar $\xrightarrow{\Delta = -\SI{11.5}{\electronvolt}}$ $\mathrm{e}^{-1}$ (\SI{8.8}{\electronvolt}) + Ar  \\   
				P7 & $\mathrm{e}^{-1}$ (\SI{13}{\electronvolt})  + Ar $\xrightarrow{\Delta = -\SI{11.5}{\electronvolt}}$ $\mathrm{e}^{-1}$ (\SI{1.5}{\electronvolt}) + Ar  \\       
    
        	\bottomrule
	    \end{tabular}
	\end{center}
    \caption{Set of inelastic processes with Argon atoms electrons can encounter in case C4}
    \label{Table_reactions_C4}	
\end{table}

In case C4, at low energies, no excitation or ionization event can occur and Coulomb collisions thermalize electrons and the EEDF follows a Maxwellian distribution, with a temperature measured as $\sim\SI{0.4}{\electronvolt}$.
It is observed that as kinetic effects become increasingly pronounced, there is a noticeable reduction in the temperature of cold electrons.
Indeed, in this regime, the beam electrons are mostly responsible for ionization and do not heat up bulk electrons.
Thus, cases C3 and C4 have a cold electron temperature of \SI{1.4}{\electronvolt} and \SI{0.4}{\electronvolt} whereas for cases with a nearly Maxwellian EEDF, such as C4 or C5 $T_e>\SI{2}{\electronvolt}$.
Consequently, estimating the ionization rate based solely on a Maxwellian distribution that depends on the measured electron temperature $T_e$ would yield highly inaccurate results in the cases of C3 and C4.

Determining whether the hollow cathode is acting in a kinetic or fluid regime depends on how rapidly electrons emitted from the cathode insert are thermalized.
For a given gas pressure and background electron density, and by taking into account the energy gain in the sheath, the collision frequency for a single electron for ionization, excitation, and Coulomb collisions can be easily computed.
Comparison of the Coulomb collision frequency with the maximum of the ionization and excitation frequencies is quantified under various conditions, as depicted in \Cref{Fig_sec4_kinetic_effects} (b).
Measured conditions, typical of plasma switch applications\citep{meshkovPlasmaBehaviorLaB62020,meshkovPerformanceLifeMeasurements2021}
, from cases C0 to C5 are presented in the figure.
It appears that whenever the EEDF deviates from a Maxwellian distribution, the ratio $\beta=\nu_{ee}/\max\{\nu_{ioniz},\nu_{excit}\}$, is significantly less than unity, which echoes previous work by Tsendin~\citep{tsendinElectronKineticsNonuniform1995,tsendinNonlocalElectronKinetics2010,tsendinElectronKineticsGlows2009}.
Indeed, the condition $\beta \ll 1$ indicates that beam electrons cannot thermalize before entering the plasma bulk, allowing their specific signature to be observed in the EEDF within the bulk plasma. In this regime, a kinetic theoretical or modeling approach should be chosen.
Conversely, whenever $\beta \gg 1$, cathode electrons are thermalized more quickly than they have to time to undergo any inelastic collision, meaning they heat up the bulk electrons and the EEDF is Maxwellian.
In this regime a fluid theoretical or modeling approach may be appropriate. 
This criterion can be succinctly expressed by,

\begin{equation}
    \beta = \frac{\nu_{ee}}{\max\left\{\nu_{ioniz}, \nu_{excit}\right\}}, \,
    \begin{cases}
        \beta \gg 1 & \Longrightarrow \text{Fluid} \\
        \beta \ll 1 & \Longrightarrow \text{Kinetic}
    \end{cases}    
    \label{Eq_beta_ratio}
\end{equation}

As an example, the C0 case presents a small deviation from a Maxwellian distribution as shown in \Cref{Fig_sec3_eedf_one_case} (a), and this can have important consequences on plasma production.
The calculation of the ionization collision frequency is given by,

\begin{equation}
    \nu_{\text{ioniz}} = n_n \int_0^{+\infty} f(E) \sigma_{\text{ioniz}}(E) \cmtextItal{v}(E) \, dE,
    \label{Eq_cal_nu_ioniz}
\end{equation}

\noindent where $f$ is the EEDF normalized to unity, $\sigma_{ioniz}$ the ionization cross-section, and $\cmtextItal{v}(E)=\sqrt{2E/m_e}$ is the electron velocity.
For cases from C0 to C5, the ionization collision frequency $\nu_{PIC}$ is computed using the EEDF $f$ obtained from the PIC simulation and presented in \Cref{Fig_sec3_eedf_one_case} (b).
These estimates are compared with the ionization collision frequency $\nu_{Maxwell}$ that is calculated assuming a hypothetical Maxwellian EEDF recalled in \Cref{Eq_eedf}.
The temperature of such an EEDF is determined by finding the best fit to the PIC data, as was done in \Cref{Fig_sec3_eedf_one_case} (a).
In case C0, according to \Cref{Eq_cal_nu_ioniz}, the PIC simulation yields an ionization collision frequency of $\nu_{PIC}\approx\SI{7.5e4}{\per\second}$.
In contrast the Maxwellian fit derived from \Cref{Fig_sec3_eedf_one_case} (a) estimates $\nu_{Maxwell}\approx\SI{1.1e+05}{\per\second}$.
Consequently, the PIC model predicts a production rate approximately 68\% lower than that estimated using a fluid approach.
For case C3, which exhibits a pronounced high-energy tail, the ratio $\nu_{PIC}/\nu_{Maxwell}\approx 35$.
This substantial discrepancy indicates a significant potential error in fluid modeling for this scenario.
Conversely, when $\beta>1$ a fluid modeling proves to be sufficient.
 For instance, in Case C5, the ratio $\nu_{PIC}/\nu_{Maxwell}$ is approximately 98\%, indicating that fluid modeling is adequate.

The proposed criterion offers a valuable method for assessing the operational regime of various hollow cathodes documented in the literature. 
For a hollow cathode employed in space propulsion with Xenon propellant \cite{goebelHollowCathodeTheory2005a}, we have a density of approximately \SI{5e20}{\per\cubic\meter} and an energy beam ranging between 10 to \SI{15}{\volt}. 
This results in a $\beta$ value of roughly \SI{15}{}, under the assumption of a pressure near \SI{1}{\torr}. 
Consequently, this suggests that a fluid treatment of electrons within the channel is appropriate, which is corroborated by extensive research \citep{mikellidesHollowCathodeTheory2005a,saryHollowCathodeModeling2017a}.
In contrast, thermionic cathodes, with applications as current and voltage stabilizers \citep{bogdanovModelingShortDc2013a,demidovGasdischargePlasmaSources2007,mustafaevSharpTransitionTwo2014}, demonstrate a different behavior. 
These cathodes exhibit dual electron populations: a cold Maxwellian group and a fast electron beam. 
In experiments involving Helium at \SI{1}{\torr}, the cold electron density was reported as \SI{5.35e18}{\per\cubic\meter}, with a beam energy of \SI{27}{\electronvolt}. 
This yields a $\beta$ value of approximately 0.03, justifying the application of a kinetic treatment.

Finally, results from this study echo several findings from the investigation from \refonlinecite{meshkovElectricalThermalCharacteristics2024}.
First, it is reported that the voltage drop in the cathode channel decreases as the current density in the external circuit increases and reaches a plateau for the tested noble gases, i.e. Argon, Helium, and Xenon. 
The plateau value converges towards the ionization or excitation potential depending on whether direct or two-step ionization is the dominant mechanism for plasma formation.
This trend is reproduced by our cases C1, C0, C2 and C5, for which $J_0$ goes from \SI{2.06}{\kilo\ampere\per\meter\squared} to \SI{80}{\kilo\ampere\per\meter\squared} and the measured potential drop ranges from \SI{16.7}{\volt} to \SI{15.8}{\volt}.
As two-step ionization is not modeled here, the potential cannot drop below the ionization potential and only direct ionization is possible. 
Besides, the authors in \refonlinecite{meshkovElectricalThermalCharacteristics2024} report that the ionization mechanism seems to vary with the plasma density.
They suggest that, at low density, beam electrons directly ionize the neutral gas whereas when the density is higher, beam electrons heat bulk electrons first.
The tail of the Maxwellian EEDF is then responsible for ionization.
Although these two mechanisms are also identified in the present paper, the governing criterion depends on the parameter $\beta$ instead of solely plasma density.
 Depending on the regime they operate, accounting for the non-Maxwellian nature of the EEDF may be necessary in the modeling of hollow cathodes for plasma switches since it is reported in \refonlinecite{meshkovElectricalThermalCharacteristics2024} that failing to do so led to systematic errors in their 0D fluid model.





\section{Analysis of the hollow cathode plume region}
\label{sec_plume_region}

In this Section we derive an analytical expression of the plasma expansion in the plume.

\begin{figure}[!htb] 
	\centering
	\includegraphics{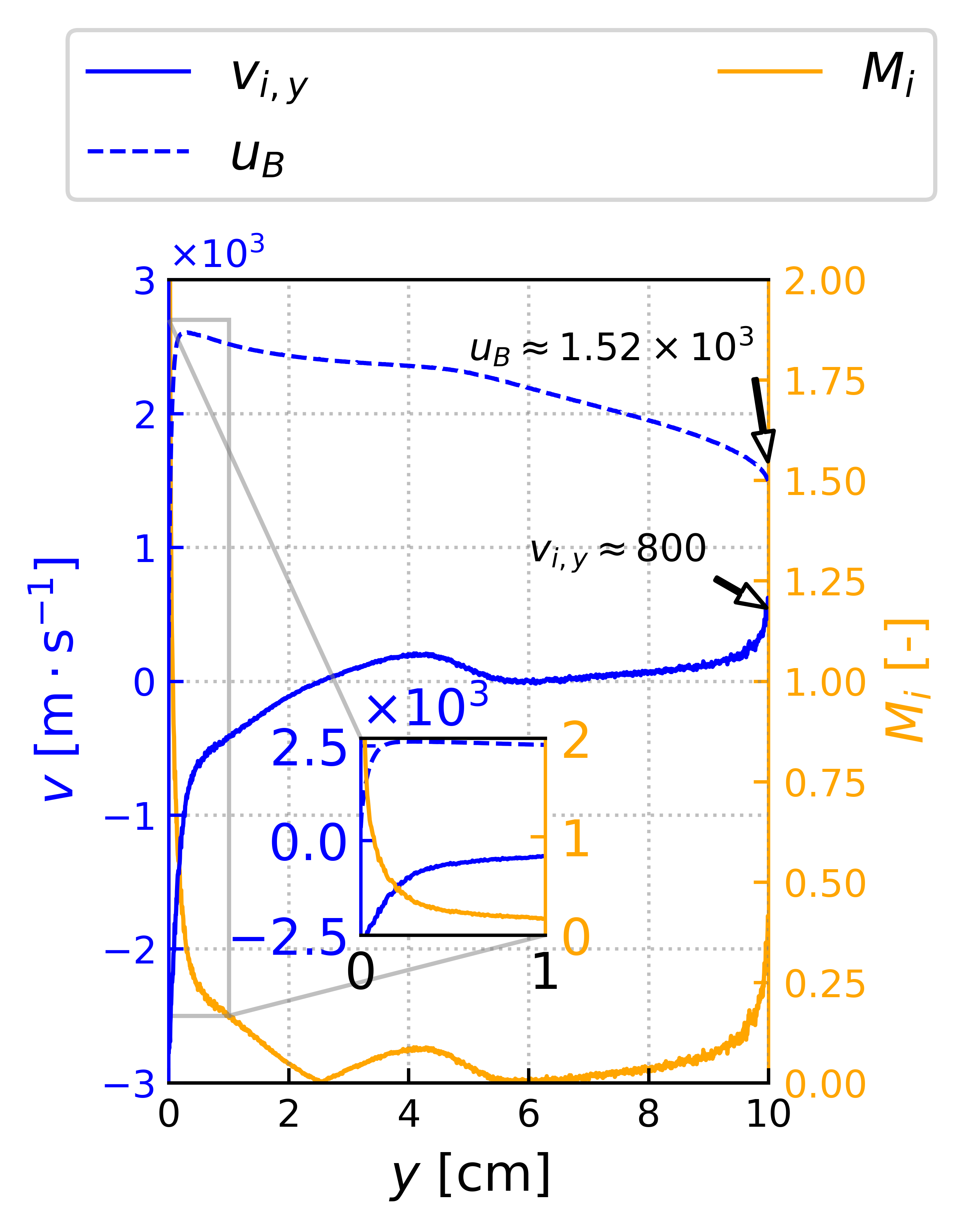}
	\caption{1D profile of the ion velocity $\cmtextItal{v}_{i,y}$, Bohm velocity $u_B$ and ion Mach number $M_i$ along the centerline of the domain at $x=0$ shown in \Cref{Fig_results_grid_plot} (e). The inset zooms in on the ion sheath at $y=0$ and reveals a supersonic speed for ions.}
	\label{Fig_velocity_bohm}
\end{figure}

In the plume region, a potential drop of $\sim$\SI{12.5}{\volt}, as shown on \Cref{Fig_results_grid_plot} (c), confines most of the cold electrons in the radial direction.
Electrons extracted from the channel escape through the anode in the axial direction, where the sheath is absent as illustrated in \Cref{Fig_1D_cuts_general} (a).
In \Cref{Fig_velocity_bohm}, profiles of the mean ion velocity $v_{i,y}$, Bohm velocity $u_B$ and the corresponding Mach number $M_i$ are shown at the centerline $x=0$ cm.
At the anode ($y=\SI{10}{\centi\meter}$), the ion velocity increases, up to $\sim$\SI{800}{\meter\per\second}, although this is insufficient to reach supersonic speeds.
The Bohm velocity is found to be $u_B \approx\sqrt{eT_e/m_i} \approx 1.5\SI{2e3}{\meter\per\second}$, as the electron temperature is locally $\sim$1-\SI{1.5}{\electronvolt}.
Absence of an anode sheath can be attributed to the presence of the external circuit in which a constant current source is imposed, indicating that quasineutrality is maintained up to the anode. 

\begin{figure}[!htb] 
	\centering
	\includegraphics{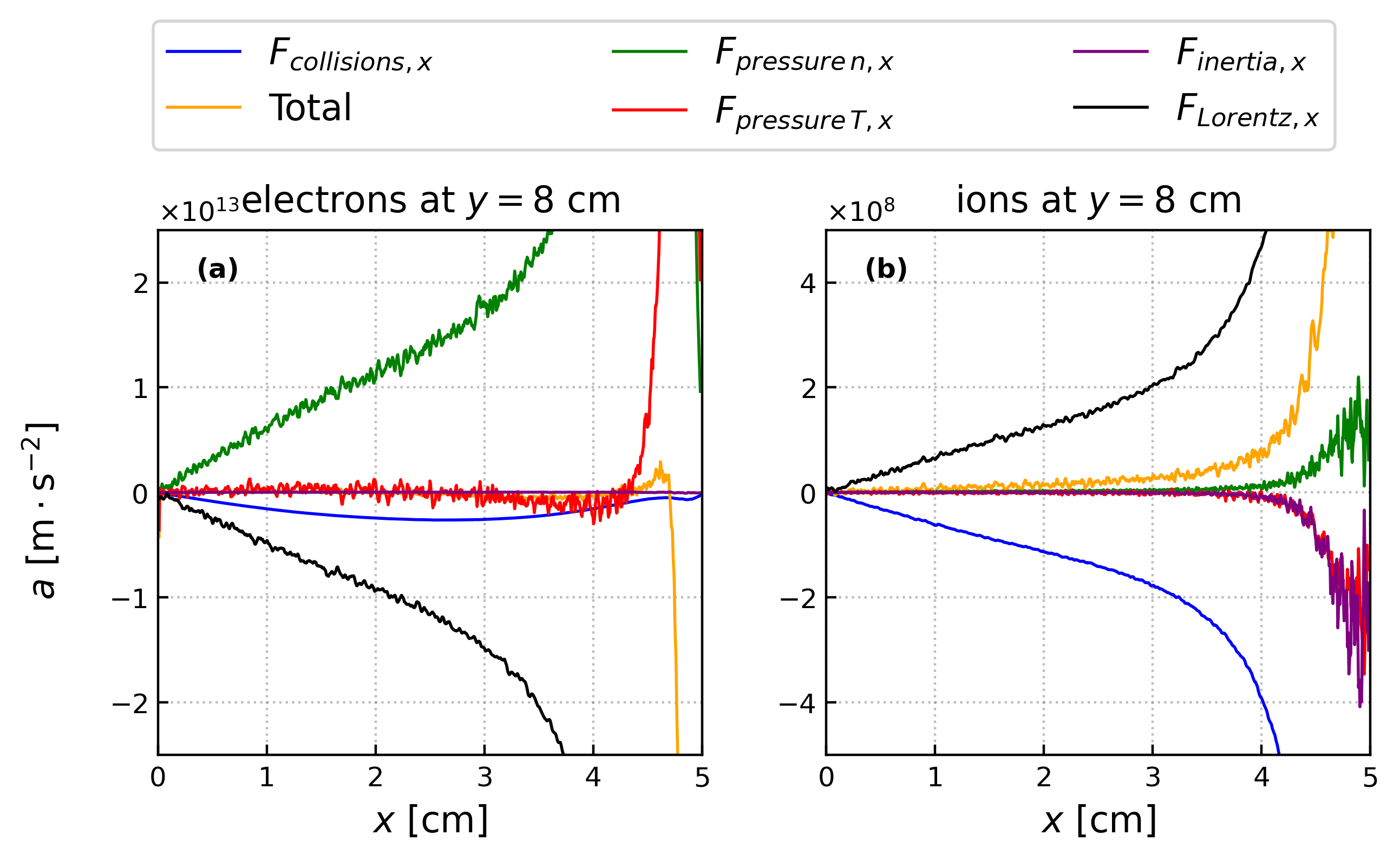}
	\caption{Momentum balance in $x$ direction at $y=\SI{8}{\centi\meter}$ for (a) electrons and (b) ions. For electrons, the density gradient, the Lorentz force and the collision terms are found to be the most important. For ions, the Lorentz force and the collision terms are predominant.}
	\label{Fig_sec3_momentum_balance_x}
\end{figure}

\begin{figure}[!htb] 
	\centering
	\includegraphics{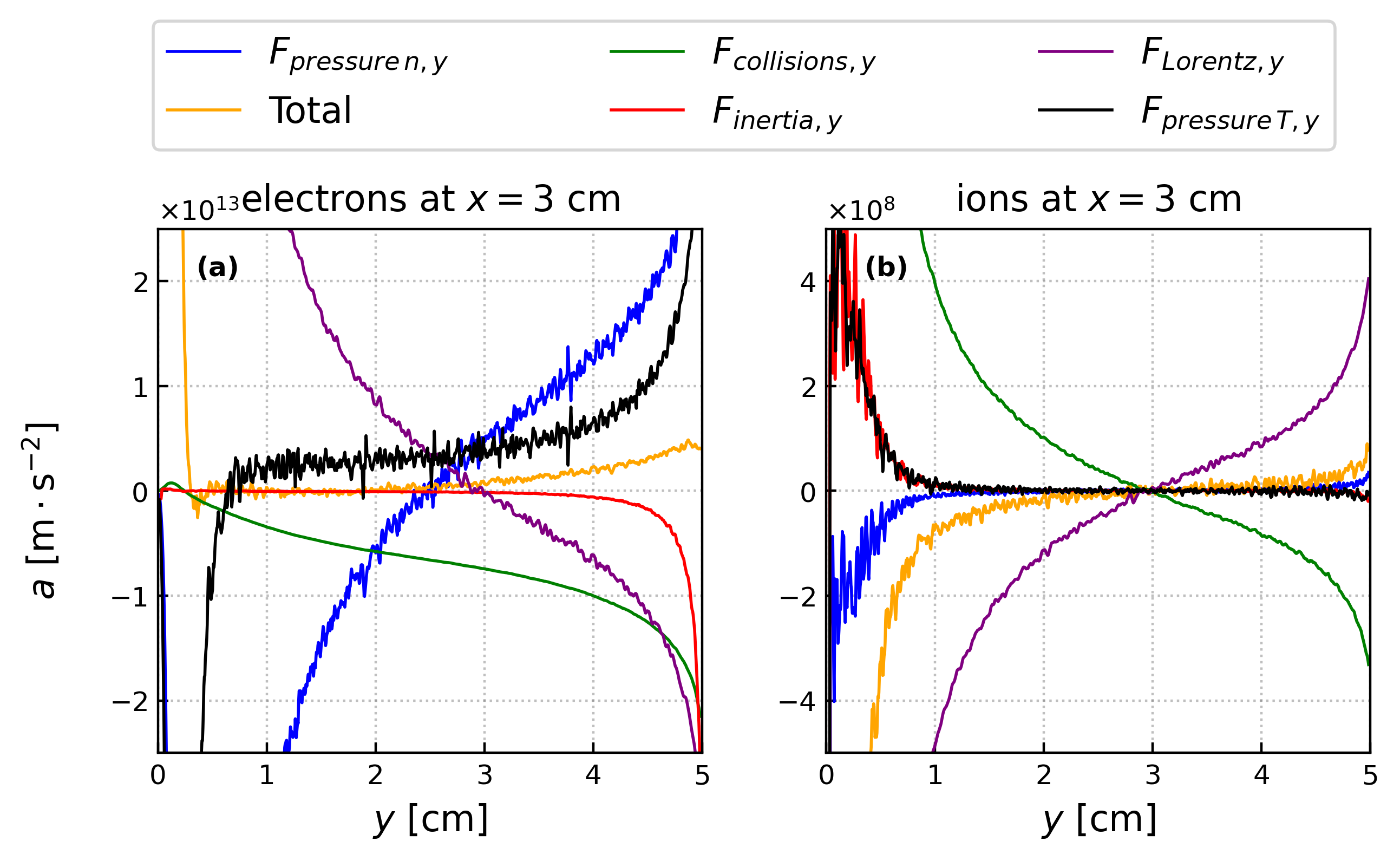}
	\caption{Momentum balance in $y$ direction at $x=\SI{3}{\centi\meter}$ for (a) electrons and (b) ions. For electrons, the density gradient, the Lorentz force and the collision terms are found to be the most important. For ions, the Lorentz force and the collision terms are predominant.}
	\label{Fig_sec3_momentum_balance_y}
\end{figure}

Understanding how the plasma expands in the plume can be better understood by evaluating the contributions of the different terms in the steady-state momentum equation,

\begin{equation}
	0=\underbrace{\vphantom{\frac{e}{m_s}}-\cmtextItal{v}_{s, x} \partial_x \cmtextItal{v}_{s, \alpha}-\cmtextItal{v}_{s, y} \partial_y \cmtextItal{v}_{s, \alpha}}_{\mathclap{F_{inertia, s,\alpha}}}
	  \underbrace{-\frac{e T_{s, y}}{n_s m_s} \partial_\alpha n_s}_{F_{pressure\,T, s,\alpha}}
	  \underbrace{-\frac{e}{m_s} \partial_\alpha T_{s, \alpha}}_{F_{pressure\,n, s,\alpha}}+
	  \underbrace{\frac{q_s}{m_s} E_\alpha}_{F_{Lorentz, s,\alpha}}
	  \underbrace{\vphantom{\frac{e}{m_s}}-\cmtextItal{v}_{s, \alpha}\nu_{tot, s} }_{\mathclap{F_{collisions, s, \alpha}}},
	\label{Eq_sec3_momentum_plume}
\end{equation}
where the index $s\in \{e,i\}$ denotes electrons and ions, and index $\alpha$ represents the $x$ or $y$ coordinate.
$q$ and $\nu_{tot}$ are the charge of the species and the total collision frequency, respectively. 
For ions, $\nu_{tot,i}$ is the charge exchange collision frequency while for electrons this encompasses elastic, excitation, and ionization collisions.
All quantities, i.e., velocity, collision frequency, temperature or electric field are self consistently obtained from the PIC simulation and \Cref{Fig_sec3_momentum_balance_x,Fig_sec3_momentum_balance_y} show the momentum balance in the $x$ and $y$ directions, respectively, for both electrons and ions. 
These figures reveal that the ion motion within the plasma bulk is mostly governed by the electric field.
As shown in \Cref{Fig_sec3_momentum_balance_x} (a) and \Cref{Fig_sec3_momentum_balance_y} (a), the electron motion is determined by competing effects between the deceleration caused by the electric field and collisions, and acceleration caused by the pressure gradient, mainly attributable to the density drop.
Based on PIC simulation data, we assume the temperature for ions and electrons to be mostly uniform in regions distant from the walls, respectively located at $y=5$, $x=5$, and $y=10$. 
Our preliminary analysis leads us to approximate the ion temperature, $T_i$, at $\sim\SI{0.02}{\electronvolt}$, and the electron temperature, $T_e$, at $\sim\SI{1.5}{\electronvolt}$.
Based on these observations, a number of small terms can be removed from the momentum balance, simplifying it to,

\begin{equation}
	\begin{gathered}
		\mathbf{v}_i=\frac{e}{m_i \nu_{i,tot}} \mathbf{E} \\
		\mathbf{v}_e=\frac{-e}{m_e \nu_{e, tot}} \mathbf{E}+\frac{-e T_e}{v_{e, tot} m_e n_e} \nabla n_e
	\end{gathered}
	\label{Eq_sec3_simplified_momentum_plume}
\end{equation}

\noindent As ionization is limited to the channel region the continuity equations for charged particle $s$ can be written as,

\begin{equation}
	\nabla \cdot\left(n_s \mathbf{v}_s\right)=0.
	\label{Eq_sec3_continuity_plume}
\end{equation}

\noindent Assuming quasineutrality $n_i\approx n_e$ and combining \Cref{Eq_sec3_simplified_momentum_plume} and \Cref{Eq_sec3_continuity_plume}, we obtain a Laplace-diffusion equation for $n_i$:

\begin{equation}
	\Delta n_i =0.
	\label{Eq_sec3_laplace_plume}
\end{equation}

\begin{figure}[!htb] 
	\centering
	\includegraphics{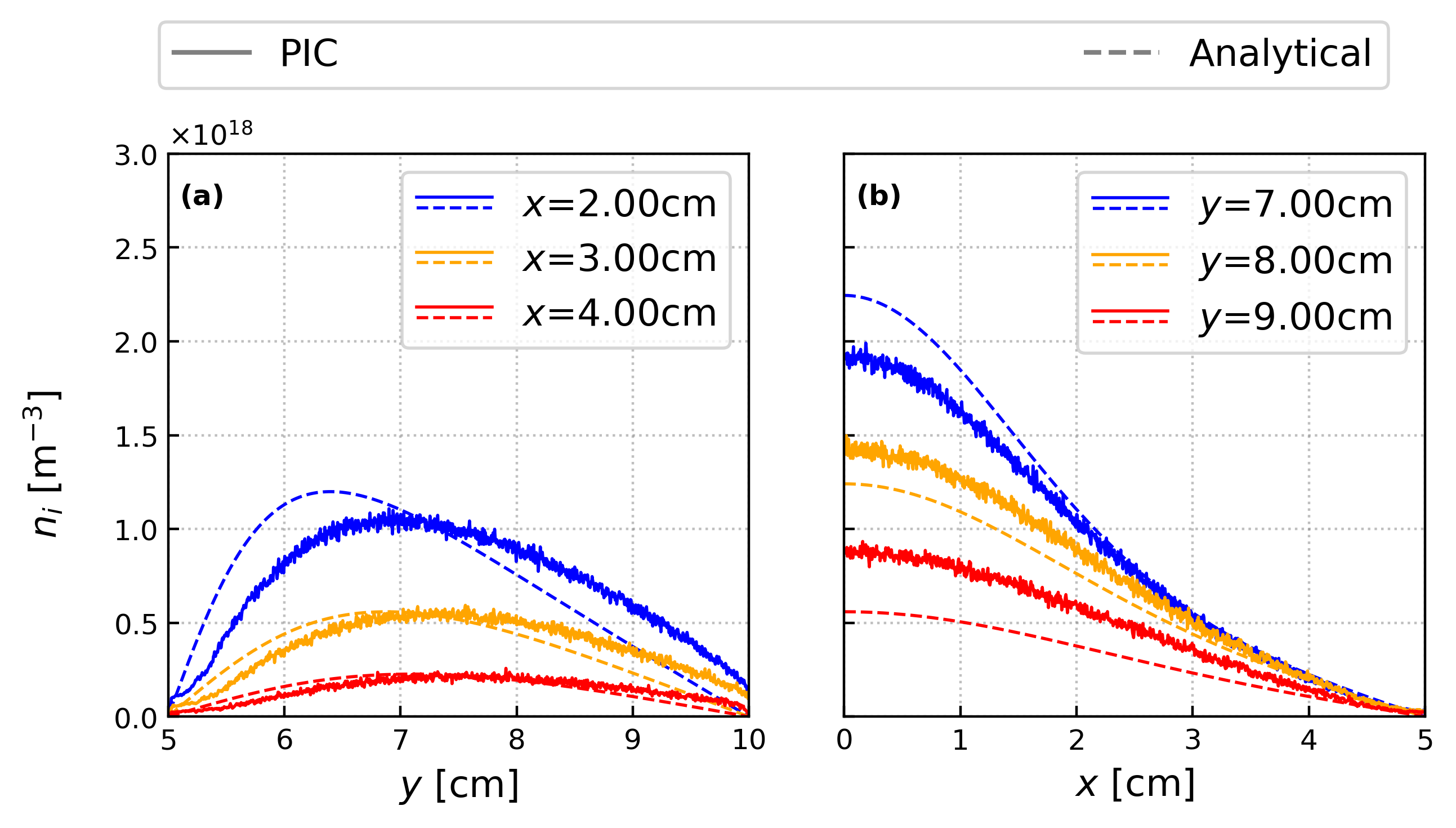}
	\caption{Steady state profiles of density $n_i$ in the $x$ direction from EDIPIC (solid line) and from the model given by \Cref{Eq_sec_3_plume_expansion} at location $y=\SI{6}{\centi\meter}$, \SI{7}{\centi\meter} and \SI{8}{\centi\meter}.}
	\label{Fig_sec3_comparison_plume_analytical}
\end{figure}

\noindent This equation is solved by separation of variables in Cartesian coordinates, with appropriate boundary conditions.
At the symmetry axis a Neuman boundary condition $\partial_x n_i = 0$ is imposed while the density must be zero at the dielectric layer.
At the top surface of the cathode we also have $n_i\approx 0$, while the density is assumed to be constant for $x \in [0,r]$ at $y=L_c$.
At the anode, the density is assumed to be much smaller than the density in the bulk and so the approximation $n_i\approx 0$ is chosen for the sake of simplicity.
The general solution can be written as,

\begin{equation}
	\begin{gathered}
		n_e=\sum_{m=0}^{\infty} \alpha_m \cos \left(k_m x\right)\left(-\tanh \left(k_m L_y\right) \cosh \left(k_m y\right)+\sinh \left(k_m y\right)\right) \\
		\alpha_m=-\frac{2}{L_x \tanh \left(k_m L_y\right)} \int_0^{L_x} n(x, 0) \cos \left(k_m x\right) \mathrm{d} x = -\frac{4n_c}{\tanh(k_mL_y)k_mL_x}\sin(k_mr),
	\end{gathered}
	\label{Eq_sec_3_plume_expansion}
\end{equation}

\noindent where $\alpha_m$ is a coefficient determined by the boundary condition $n_i(x,L_y)\approx 0$ and $n_c$ is the plasma density in the channel at $y=5$, either measured from the PIC simulation or obtained by estimates in the channel. 
In cylindrical coordinates, the plasma plume expansion would involve Bessel functions of the first kind $J_m$ instead of sinusoidal functions.
Any profile of $n_c$ can be used but in this case we choose to select a constant value for simplicity. 
In the present estimate, the value of $n_c = \SI{9e18}{\per\cubic\meter}$, corresponding to the maximum density found in the channel is used instead of $\sim\SI{1e18}{\per\cubic\meter}$, which is the approximate average value measured at the channel exit ($y=\SI{5}{\centi\meter}$).
The model described by \Cref{Eq_sec_3_plume_expansion} assumes the electron temperature to be uniform, particularly in the $y$ direction.
However a closer examination of \Cref{Fig_sec3_momentum_balance_y} (a) shows that the pressure gradient due to a slight temperature drop in the plume is present and accelerates electrons.
Thus, in the PIC simulation, this additional contribution tends to extract more electrons from the cathode and so the plasma density in the plume is therefore expected to be slightly higher than the model estimate.
Adjusting the parameter $n_c$ to a higher value, we obtain density magnitudes that are consistent with the PIC data, as illustrated in \Cref{Fig_sec3_comparison_plume_analytical} panels (a) and (b).
In \Cref{Fig_sec3_comparison_plume_analytical} (b), with this adjustment the density profile in the $x$ direction is slightly overpredicted at $y=7$ cm.
This could be attributed to the fact that a uniform density is assumed at the channel exit ($y=5$ cm) whereas the density drops to 0 near at the upper left corner of the cathode as shown in \Cref{Fig_results_grid_plot} (a).
At $y=8$ cm and $y=9$ cm in \Cref{Fig_sec3_comparison_plume_analytical} (b) density profiles are slightly underestimated most likely due to the aforementioned uniform temperature assumption.
Despite these discrepancies, the overall profile and trend in both the $x$ and $y$ directions is reasonably predicted, which is a crucial input for hollow cathodes designed for plasma switch applications. 
This tool serves as an efficient means to estimate plasma uniformity within the plume and adjacent to the external circuit.
The plasma uniformity predicted by \Cref{Eq_sec_3_plume_expansion} can be estimated by the standard deviation of a 1D profile in the $x$ direction at a given $y$ location.
The standard deviation decreases significantly at a distance in the $y$ direction that is at least equal to half of the radial length, denoted as $\Delta y = \frac{L_x}{2}$.
In the present case, where the plume length in the $y$ direction is equal to $L_x$, the standard deviation decreases by more than 50\% compared to its value at the channel exit at $y=\SI{7.5}{\centi\meter}$.
Therefore, plasma uniformity is likely to improve when the plume length-to-width ratio is at least greater than unity.

This study did not detect the presence of high-energy ions. 
However, prior investigations\citep{haraIonKineticsNonlinear2019,jornsIonAcousticTurbulence2014} indicate that in electric propulsion systems, high-energy ions may arise in the plume when the electron temperature $T_e$ is significantly higher than the ion temperature $T_i$, coupled with an electron Mach number that notably exceeds unity.
The underlying mechanism for this phenomenon is attributed to ion acoustic turbulence (IAT) that appears under these conditions. 
In the present application, $T_e \gg T_i$, however the electron drift reaches a maximum of $\cmtextItal{v}_e \sim \SI{e5}{\meter\per\second}$ and when compared to a thermal speed of $\cmtextItal{v}_{e,th} \sim \SI{8e5}{\meter\per\second}$, this indicates that the IAT does not arise.
In hollow cathodes for space propulsion, the neutral gas pressure dramatically drops in the plume, below \SI{1}{\milli\torr}, making the plasma virtually collision-free and allowing electrons to reach a supersonic regime. 
In another study, authors in \refonlinecite{hallEffectFacilityBackground2021} found that increasing the background pressure from 20 to \SI{100}{\micro\torr} was enough to reduce the energy of high-energy ions by 35\%. However with our gas pressures approximately 100 times higher, it is unsurprising that fast ions do not exist.


\section{Conclusion}

In this paper, 2D3V PIC simulations of an argon hollow cathode for plasma switch application have been performed.
Analysis of the energy distribution functions for both ions and electrons revealed that non-Maxwellian behavior can occur in the channel region.
For electrons, it was found that a Maxwellian EEDF is present only when the thermalization of electrons occurs sufficiently fast.
This happens if Coulomb collisions are frequent enough to transfer the energy from the beam electrons to the cold electron population before they have time to ionize or excite neutral particles.
The ratio $\beta$, as defined in \Cref{Eq_beta_ratio}, compares the electron-electron collision frequency with the maximum collision frequency between direct ionization and excitation for beam electrons.
It serves as a straightforward criterion to determine whether fluid or kinetic modeling approaches are suitable for hollow cathode simulations. 
Whenever $\beta\gg 1$, then beam electrons quickly thermalize and a fluid model is sufficient to describe the system.
In contrast, if $\beta\ll 1$, then beam electrons play an essential role in ionization processes and the EEDF in the plasma bulk is not Maxwellian, requiring a kinetic treatment. 
We found that for most hollow cathodes used in space propulsion, a fluid treatment of electrons in the channel is a reasonable assumption, which is reflected by state-of-the-art models \citep{saryHollowCathodeModeling2017a,mikellidesNumericalSimulationsPartially2015}.
In contrast, some applications related to plasma deposition processes \cite{borisHollowCathodeEnhanced2022} or to a plasma switch \citep{meshkovElectricalThermalCharacteristics2024} may operate in regimes for which a kinetic treatment is required. 
Finally, in the case of plasma switch application, a momentum balance showed that the ion dynamics is mostly governed by the local electric field.
The electron motion is the result of a balance between a decelerating electric field, and an accelerating density gradient.
Overall, the plasma expansion in the plume is mostly determined by diffusion allowing for the development of a simple analytical model sufficient to estimate the plasma uniformity in the plume. 

To further our understanding of the ionization mechanism, particularly within hollow cathodes utilized in plasma switch applications, several follow-up studies are possible.
First, modeling two-step ionization processes by tracking metastable species would reveal under which conditions they become more frequent than direct ionization.
Refining such prediction could be a useful input for future design as it was suggested in \refonlinecite{meshkovElectricalThermalCharacteristics2024} that metastables may have an effect on the potential drop in these devices. 
Finally, in order to help the design of future devices, it would certainly be beneficial to account for a self consistent thermionic electron injection scheme, one which depends on the cathode temperature, and accounts for self-induced magnetic field generated by high currents.

\begin{acknowledgments}
The information, data, or work presented herein was funded in part by the Advanced Research Projects Agency-Energy (ARPA-E), U.S. Department of Energy, under Award Number DE-AR0001107.
The views and opinions of authors expressed herein do not necessarily state or reflect those of the United States Government or any agency thereof.
The authors thank Dr. David J. Smith and Dr. Svetlana E. Selezneva for fruitful discussions.
\end{acknowledgments}

\section*{DATA AVAILABILITY}
The data that support the findings of this study are available from the corresponding author upon reasonable request.
\bibliography{central_bibtex_library_zotero}

\end{document}

%% file: customed_settings_pop.tex
\usepackage{graphicx}
\usepackage{dcolumn}
\usepackage{bm}

\usepackage[utf8]{inputenc}
\usepackage[T1]{fontenc}

\usepackage{mathptmx}
\newcommand{\cmtextItal}[1]{\textit{\text{\fontfamily{cmr}\selectfont #1}}}
\usepackage{etoolbox}
\usepackage{soul}

\usepackage{placeins} 

\usepackage[USenglish]{babel}
\usepackage{physics}
\usepackage{cleveref}

\usepackage[binary-units]{siunitx}
\DeclareSIUnit\gauss{G}
\DeclareSIUnit\torr{Torr}
\usepackage{todonotes}

\usepackage{mathtools}

\usepackage{booktabs}
\newcolumntype{C}{>{$}c<{$}}
\newcolumntype{L}{>{$}l<{$}}
\newcolumntype{R}{>{$}r<{$}}
\newcommand{\ra}[1]{\renewcommand{\arraystretch}{#1}}
\newcommand{\specialcellLeft}[2][c]{%
  \begin{tabular}[#1]{@{}l@{}}#2\end{tabular}} 

\newcommand{\refonlinecite}[1]{Ref.~[\onlinecite{#1}]}
\newcommand{\refsonlinecite}[1]{Refs.~[\onlinecite{#1}]}

